\documentclass[12pt,preprint]{emulateapj}
\usepackage{lscape,url,color}
\usepackage{amsmath}
\usepackage{relsize}

\shorttitle{Indications of Water Clouds in W0855}
\shortauthors{Faherty et al.}

\begin{document}

\title{Indications of Water Clouds in the Coldest Known Brown Dwarf\footnote{This paper includes data gathered with the 6.5 meter Magellan Telescopes located at Las Campanas Observatory, Chile.}}

\author{Jacqueline K.\ Faherty\altaffilmark{1,2,3}, C.G. Tinney\altaffilmark{4,5}, Andrew Skemer\altaffilmark{6}, Andrew J. Monson\altaffilmark{7}}

\altaffiltext{1}{Department of Terrestrial Magnetism, Carnegie Institution of Washington, Washington, DC 20015, USA; jfaherty@ciw.edu }
\altaffiltext{2}{Department of Astrophysics, 
American Museum of Natural History, Central Park West at 79th Street, New York, NY 10034}
\altaffiltext{3}{Hubble Fellow}
\altaffiltext{4}{ School of Phsyics, UNSW Australia. 2052. Australia}
\altaffiltext{5}{ Australian Centre for Astrobiology, UNSW Australia. 2052. Australia}
\altaffiltext{6}{Steward Observatory, Department of Astronomy, University of Arizona,Tucson, AZ 85721, USA  }
\altaffiltext{7}{Observatories of the Carnegie Institution of Washington, Pasadena, CA 91101, USA }

\begin{abstract}
We present a deep near-infrared image of the newly discovered brown dwarf WISE J085510.83-071442.5  (W0855) using the FourStar imager at Las Campanas Observatory.  Our detection of $J3$=24.8$^{+0.53}_{-0.35}$ (J$_{MKO}$=25.0$^{+0.53}_{-0.35}$) at 2.6$\sigma$ -- or equivalently an upper limit of  $J3$ $>$ 23.8 (J$_{MKO}$ $>$ 24.0) at 5$\sigma$ makes W0855 the reddest brown dwarf ever categorized (J$_{MKO}$ $-$ W2 = 10.984$^{+0.53}_{-0.35}$ at 2.6$\sigma$ -- or equivalently an upper limit of J$_{MKO}$ $-$ W2 $>$ 9.984 at 5$\sigma$) and refines its position on color magnitude diagrams.  Comparing the new photometry with chemical equilibrium model atmosphere predictions, we demonstrate that W0855 is 2.7$\sigma$ from models using a cloudless atmosphere and well reproduced by partly cloudy models (50\%) containing sulfide and water ice clouds.   Non-equilibrium chemistry or non-solar metallicity may change predictions, however using currently available model approaches, this is the first candidate outside our own solar system to have direct evidence for water clouds.  
\end{abstract}


\section{INTRODUCTION}
Brown dwarfs are substellar mass objects with effective temperatures (T$_{eff}$) ranging from stellar-like at the high end ($\sim$ 3000 K) to planet-like at the low-end.  Many brown dwarfs are found in isolation (making direct observations and spectra obtainable), and they can share T$_{eff}$s, luminosities, ages, and/or masses with directly imaged exoplanets. Therefore they are a gateway population to characterization studies (e.g. \citealt{Delorme12}, \citealt{Faherty13,Faherty13a}, \citealt{Kirkpatrick12}, \citealt{Liu13}, \citealt{Beichman14}).  

NASA's Wide-Field Infrared Explorer (WISE) mission launched in late 2009, revolutionized our understanding of the coldest brown dwarfs (\citealt{Wright10}). Follow-up of WISE selected targets led to the discovery of nearly 20 objects in the T$_{eff}$$\sim$300-500 K temperature range and the definition of a new spectral subtype -- the ``Y" dwarfs (\citealt{Cushing11,Cushing14}; \citealt{Kirkpatrick11,Kirkpatrick12,Kirkpatrick13}; \citealt{Tinney12}; \citealt{Burgasser11}).  These cold brown dwarfs have distinctly lower luminosity (\citealt{Dupuy13}, \citealt{Tinney14})  from warmer ``T" dwarf sources that have been studied for the past 15 years  (e.g. \citealt{Lucas11},\citealt{Burningham11,Burningham11a}).    

Y dwarf temperatures are more similar to those of the cold, gas-giant exoplanets known to orbit solar-aged stars than are warmer brown dwarfs: the temperatures and atmospheres of which have more in common with low-mass stars. Their spectra are characterized by remarkably strong CH4 absorption, resulting in J and H band peaks that are narrower than those for warmer T dwarfs (\citealt{Cushing11}).  Studies of the population of these extremely cold brown dwarfs leads to a measurement of both the ability of star formation processes to make brown dwarfs at the lowest masses, and the mass cut-off for the star formation process itself.  Studies of their individual spectral energy distributions reveal the complex chemical processes that dominate low-temperature atmospheres, directly applicable to interpreting exoplanet data.

Recently, \citet{Luhman14b} discovered the coldest brown dwarf yet detected (T$_{eff}$ $\sim$ 225 - 260 K) as the fourth closest system to the Sun.  The source, designated WISE J085510.83-071442.5 (here-after W0855), is important for studies of cold, gas-giant-planet-like atmospheres.  Notably, it allows us to characterize an important opacity source for studies of exoplanets -- the clouds -- at even lower temperatures than possible in work to date. 

In this paper, we present a deep $J3$ band image of W0855.  A comparison of our photometry with recent model atmosphere predictions from \citet{Morley14} provides evidence for sulfide and water ice clouds.  Section 2 details the background data on W0855. Section 3 describes the data acquired for this paper and Section 4 investigates the position of W0855 on color-magnitude diagrams.  We present the conclusions in Section 5.

\begin{figure*}[t!]
\resizebox{\hsize}{!}{\includegraphics[clip=true]{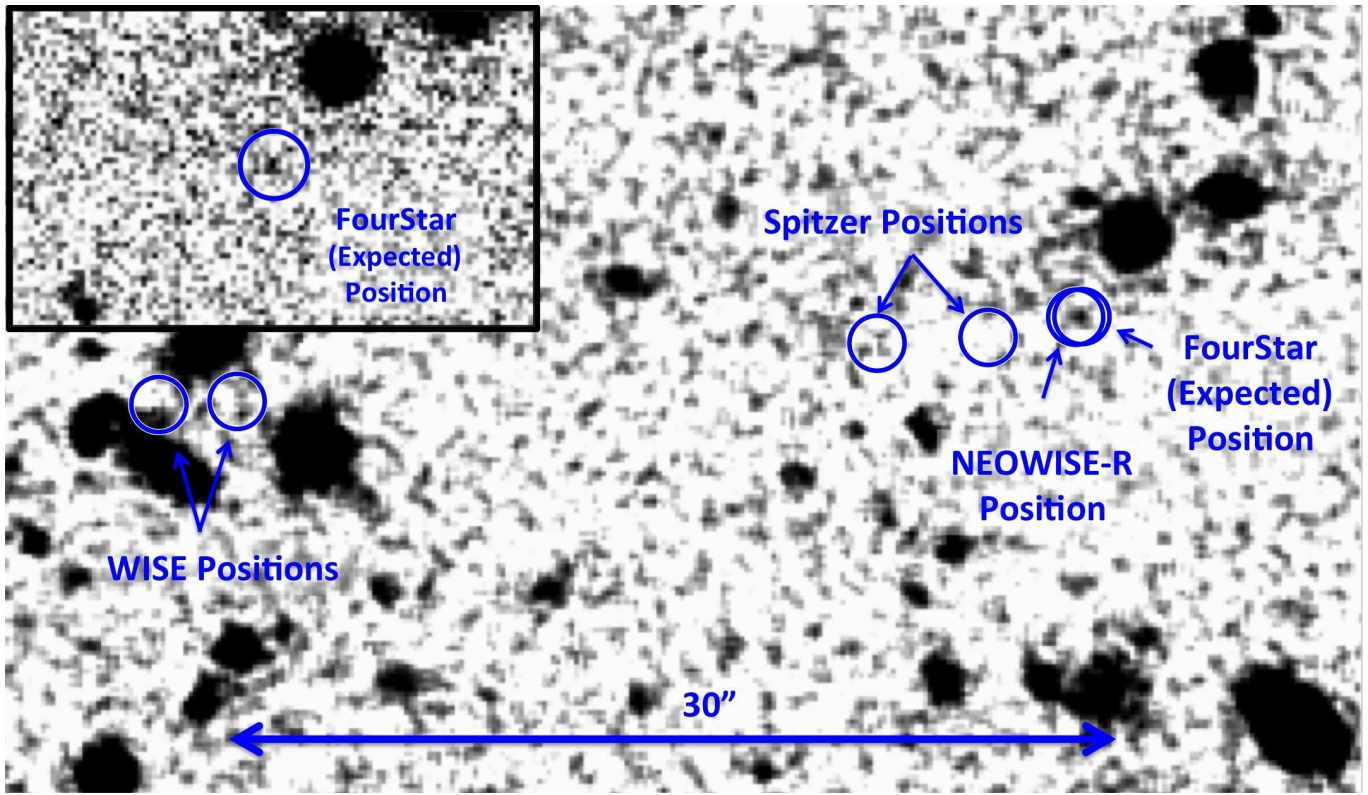}}
\caption{\footnotesize A $\sim$ 45$\arcsec$ x 25$\arcsec$ region of the final mosaic image at $J3$ created from three nights of observing the source W0855 with the FourStar imager.  For display purposes, we have gaussian smoothed the image with a kernel radius of 2.  The inset shows the location in the unsmoothed mosaic.  The total accumulated time on this image is 3.5 hr with an estimated seeing of 0.5$\arcsec$.  The WISE W2 only positions reported in  \citet{Wright14} are shown as well as the $Spitzer$ positions reported in \citet{Luhman14b} , the NEOWISE-R position from \citet{Wright14} (left), and our expected position on 2014 14 May (right).   We place a 5$\sigma$ limit of $J$3 $>$ 23.8 (J$_{MKO}$ $>$ 24.0) for the image and a 2.6$\sigma$ detection of $J3$=24.8$^{+0.53}_{-0.35}$ (J$_{MKO}$=25.0 $^{+0.53}_{-0.35}$) at the position of W0855.
}
\label{T0855_J3}
\end{figure*}

\section{Background on W0855}
In \citet{Luhman14a} and \citet{Kirkpatrick14} the high proper motion of the source W0855 was noted between the 2010 May and November WISE images.  However at the time of the original detection by WISE, it was not realized that the photometry was contaminated by a cluster of background sources which skewed the photometry and astrometry (\citealt{Luhman14b}, \citealt{Wright14}; and see the area surrounding the ``WISE Positions" on Figure~\ref{T0855_J3}).  \citet{Luhman14b} obtained $Spitzer$ $Space$ $Telescope$ $[3.6]$ and $[4.5]$$\mu$m photometry for W0855 and verified the motion.  The astrometry placed W0855 among the four closest systems to the Sun with a parallax of 0.454$\pm$0.045\arcsec~(or $\sim$2.2pc) and a proper motion of 8.1$\pm$0.1\arcsec yr$^{-1}$ (the third largest proper motion ever detected). 

Using the absolute magnitude at 4.5$\mu$m and the (J - [4.5]) color limit, \citet{Luhman14b} compared to model predictions and estimated the T$_{eff}$ for W0855 to be 225 - 260 K, making this the coldest compact source ever detected outside our own solar system. 

 \cite{Wright14} recently reported new, uncontaminated WISE photometry with NEOWISE-R\footnote{NEOWISE-R is a reactivation of WISE to search for near Earth objects.} that is in far better agreement with the expected ``Y" spectral type for this source than the first detections indicated.  \citet{Wright14} also refined the parallax and proper motion of W0855 using a new epoch, reporting $\pi$=0.448$\pm$ 0.033$\arcsec$ and $\mu$= 8.08$\pm$0.05\arcsec yr$^{-1}$.  

Over a given age range of 1 - 10 Gyr (covering all possible ages for a thin/thick disk Milky Way object), the mass of this ``brown dwarf" is 3 - 10 M$_{Jup}$ making it a truly remarkable bridge between the lowest mass objects formed by star formation processes and the giant planets we are only beginning to obtain directly imaged information on.  Indeed, it easily crosses into the ``planetary mass regime"  classically placed at 13 M$_{Jup}$ (\citealt{Burrows97}; see discussion in \citealt{Luhman14b}).

\begin{figure*}[!ht]
\begin{center}$
\begin{array}{cc}
\includegraphics[width=3.5in]{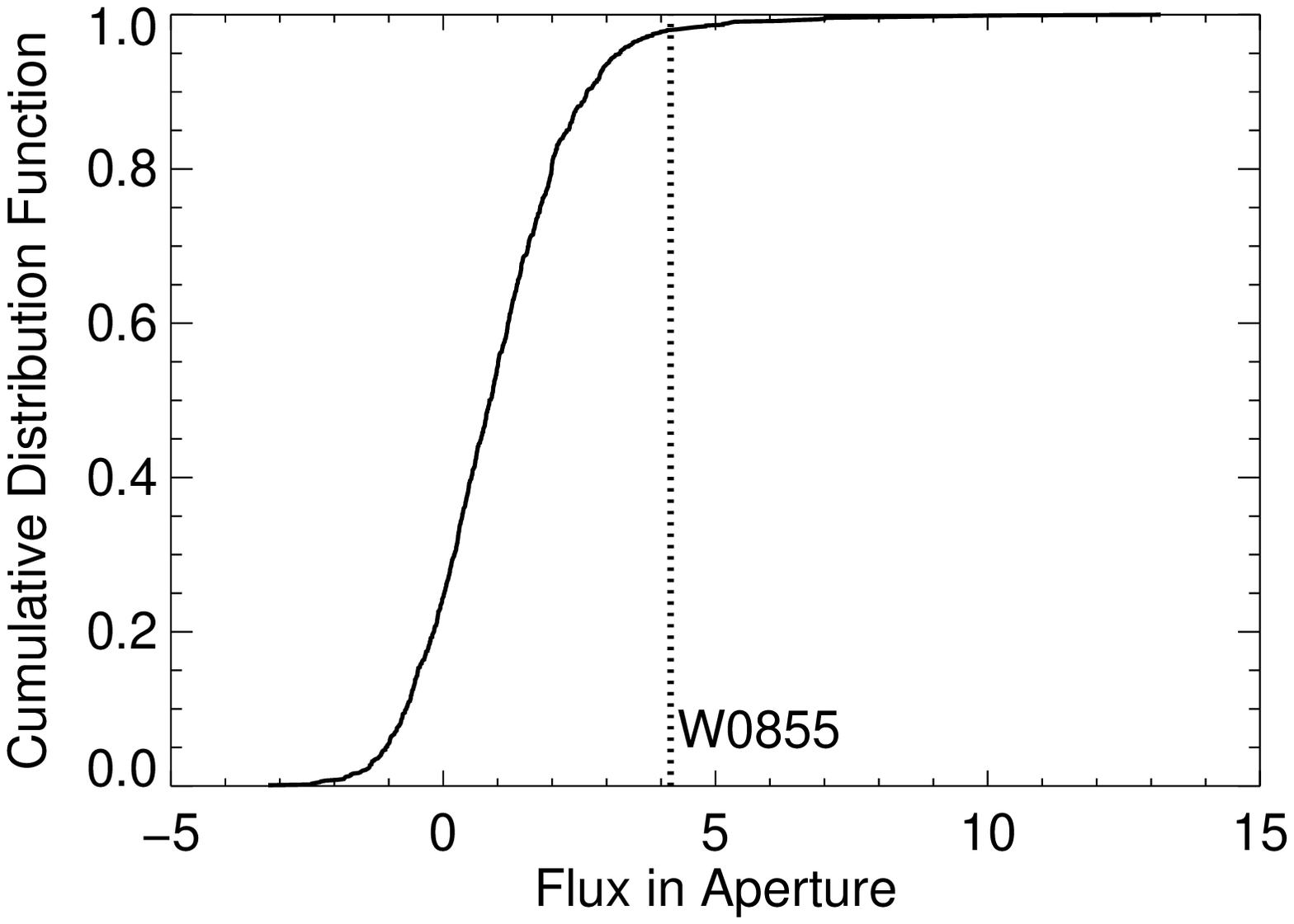}&
\includegraphics[width=3.5in]{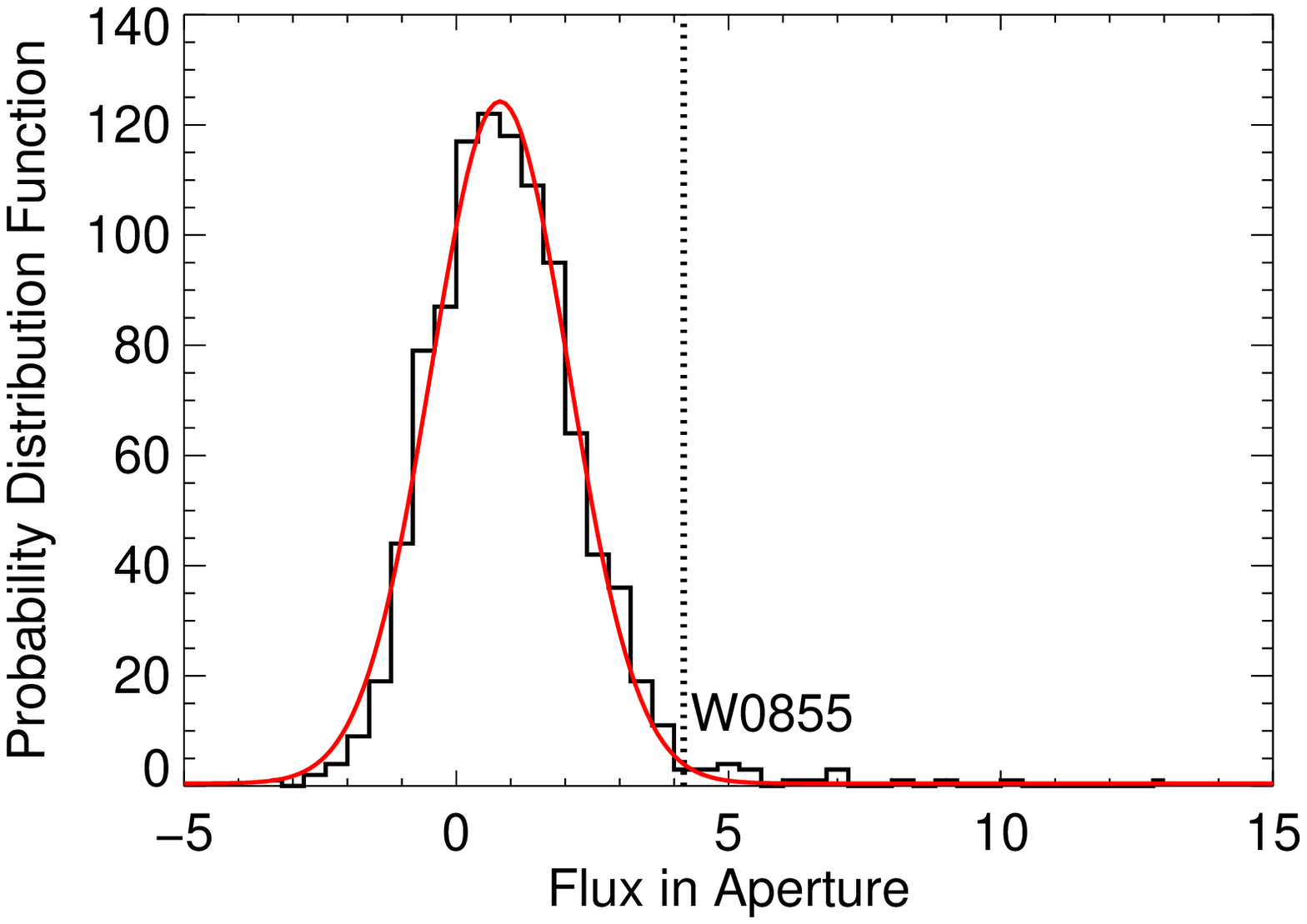}  \\

\end{array}$
\end{center}
\caption{The cumulative (left) and probability (right) distribution functions for the detection of W0855.  Plotted is the aperture photometry of W0855 and 1000 random places on our masked image (where all sources with S/N $>$ 5$\sigma$ have been masked) demonstrating that 98\% of sources are below the S/N of what we find at the expected position of the new brown dwarf.  Over-plotted in a red solid line on the probability distribution function (binned to 0.4) is a gaussian fit to the data.  We find that the detection of W0855 is valid at the 2.6$\sigma$ confidence level.  }
\label{Fig:CDF}
\end{figure*}

\section{DATA}
The majority of brown dwarfs are characterized in the near-infrared and all Y dwarfs have been classified using near-infrared spectroscopy.   As follow-up spectroscopy for W0855 is extremely difficult given its cold temperature and low luminosity, near-infrared photometry is required to place it in context with the known brown dwarf population.  Moreover, \citet{Morley14} predict that water condenses in the upper atmospheres of objects with temperatures $<$ 450 K, leading to prominent features notable in the $J$ band.  As W0855 only had a lower limit on the $J$ band photometry from \citet{Luhman14b}, there was excellent motivation for obtaining a deep image of W0855 in the near-infrared.

\subsection{FourStar Image}
W0855 was observed with the FourStar infrared mosaic camera mounted on the 6.5m Magellan Baade telescope at Las Campanas Observatory, Chile (\citealt{Persson13}) as part of an ongoing parallax program (e.g. \citealt{Tinney12, Tinney14}, \citealt{Kirkpatrick13}).   The instrument uses four 2048 x 2048 Teledyne HAWAII-2RG arrays that produce a 10.9$\arcmin$ x 10.9$\arcmin$ field of view at a plate scale of 0.159$\arcsec$ pixel$^{-1}$.  

W0855 was observed at the start of each night on 2014 May 12- 14 (UT) for 2 - 3 hours with the narrowband $J3$ filter.  As demonstrated in \citet{Tinney12}, the $J3$ filter encompasses the $J$-band opacity hole for Y dwarfs. At the same time, its narrow wavelength coverage minimizes signal from the sky, making it an optimal choice for imaging the coldest brown dwarfs. 

Total integration times collected during the imaging sequence for W0855 varied as did the seeing.  As we were observing a target at 08 hours and -07 degrees from Chile in May, we only had the start of the evening before the source fell to an unreasonably high airmass.  The first night had the poorest conditions with an average seeing of 0.9 - 1.0$\arcsec$.  The second and third nights varied but shared an average seeing of 0.4 - 0.6$\arcsec$.  

Our imaging strategy employed a 20s exposure at $J3$ with 6 co-adds and an 11-point dither pattern randomly distributed within a  15$\arcsec$ box.  We collected 50 images on the first night, 61 images on the second, and 40 images on the third. Data reduction was done using the FourStar pipeline (Monson et al. in prep).  Raw frames were linearized, dark subtracted and flat-fielded for
each of the four arrays.  Then, a first-pass local background was created for each frame using the adjacent nine in the following manner:  (1) Each nearby frame had a low-order background term fit and subtracted to
remove gradients, then the resultant images were averaged with 50\% of
the high values clipped to remove sources.  (2)  The backgrounds which
were removed were averaged together without clipping but weighted
by the differential time from the desired frame.   (3) The two backgrounds
(low-order and high-order) were added together to construct the
first-pass background for each frame.

The background subtracted images were source-extracted with the Astromatic.net program $Sextractor$,
fit with a World Coordinate System (WCS) grounded with matches to the Two Micron All-Sky Survey (2MASS; \citealt{Cutri03})  using $Scamp$ and then re-sampled using $Swarp$
to construct a first-pass mosaic image.   Point sources were then detected and individual masks updated
for each original input frame.

A second-pass sky subtraction step proceeded like the first except
instead of clipping the upper 50\% of pixel values, objects were
explicitly masked from each frame.  A final mosaic was created from all three nights by weighing each exposure by:
\begin{equation}
      weight~=~  \frac{1e7}{(fwhm^{2}~\times~stdev^{2}~\times flxscale)}
\end{equation}
where 1e7 is an arbitrary constant, fwhm is the seeing in arcseconds, stdev is the standard deviation from the background counts for each frame, and flxscale is a differential zero point term that takes into account slight airmass or weather related changes in image depth. 

The total exposure time accumulated over the three nights was 5.03 hr, however the weighting strategy resulted in a final mosaic at an exposure time of 3.5 hr with a seeing of 0.5$\arcsec$.  Consequently nearly all of the first night's data did not contribute to the final mosaic due to the poor conditions.  

To calibrate the photometry in our image, we followed the same prescription described in \citet{Tinney12}.  Within the final image, we reached a 5$\sigma$ aperture limit at $J3$ of $\sim$ 23.8 or J$_{MKO}$ $\sim$ 24.0 when the photometric relations from \citet{Tinney14} are applied (J$_{MKO}$ - J3 = 0.20$\pm$0.03 for Y dwarfs). Figure ~\ref{T0855_J3} shows an $\sim$ 45$\arcsec$  x 25$\arcsec$  region of the final mosaic containing the previous WISE positions, the \citet{Luhman14b} $Spitzer$ positions, the NEOWISE-R position from \citet{Wright14} and our expected 2014 14 May position.  

\begin{figure*}[t!]
\resizebox{\hsize}{!}{\includegraphics[clip=true]{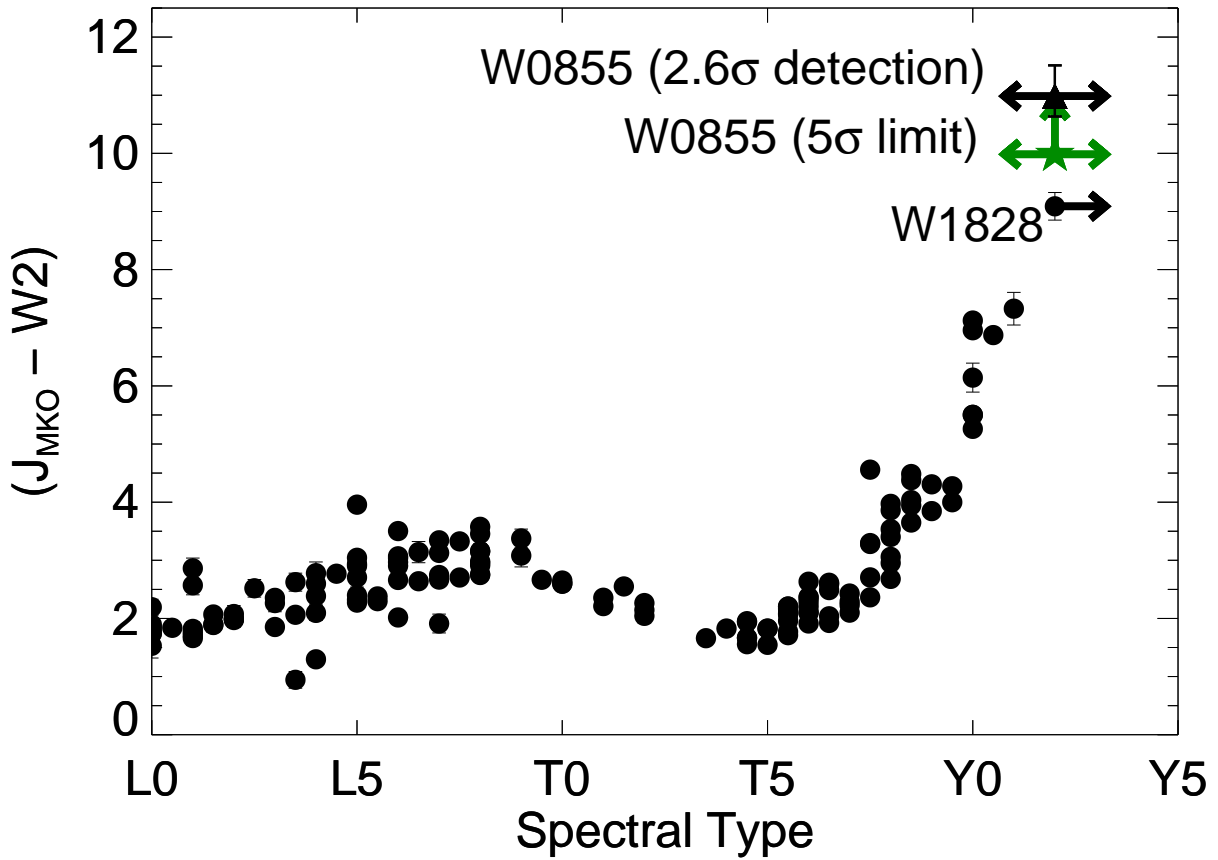}}
\caption{\footnotesize The J$_{MKO}$ $-$ W2 color versus spectral type for L, T, and Y dwarfs.  We highlight the 5$\sigma$ limit as a green five point star with upward facing arrow and the 2.6$\sigma$ detection as a filled triangle.  The x-position of W0855 is placed at a spectral type of Y2 (the same as W1828) although with the lack of spectra, this is very uncertain.}
\label{T0855_SpT}
\end{figure*}

\subsection{The 2.6$\sigma$ detection}
We investigated the expected position of W0855 in our images with great care to give a thorough limit on the detection.  For this analysis we focused only on a $\sim$ 3.0$\arcmin$ square portion of the final mosaic centered on W0855.  We first found all sources in a gaussian smoothed image (with $\sigma$ equal to the FWHM) with a signal to noise (S/N) $>$ 5$\sigma$.  We then masked those out using a radius set by where a gaussian would fall below 10\% of the noise floor.  We then conducted aperture photometry on the unsmoothed, expected position of W0855 as well as 1000 random places in the image (avoiding the masks) using a radius of 0.6731 x FWHM (Full Width at Half Maximum) -- based on the optimum size for sources whose Point Response Function (PRF) profiles can be approximated as Gaussian (Masci 2008\footnote{web.ipac.caltech.edu/staff/ fmasci/home/mystats/ GaussApRadius.pdf}).

\begin{figure*}[t!]
\resizebox{\hsize}{!}{\includegraphics[clip=true]{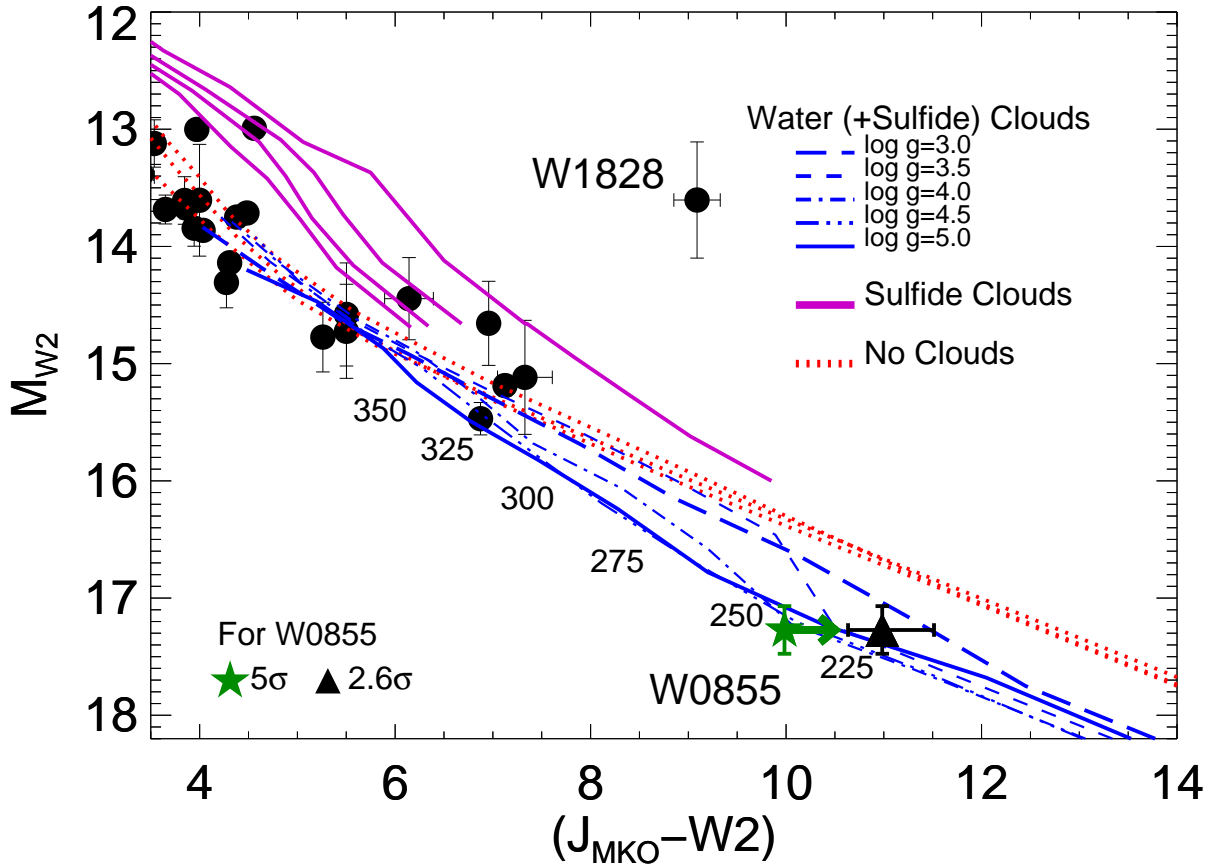}}
\caption{\footnotesize J$_{MKO}$ $-$ W2 color versus absolute magnitude at $W2$ for objects with parallaxes and spectral types later than T8.5.   Over-plotted as purple solid lines are the sulfide cloud model predictions from \citet{Morley12} with the sedimentation parameter (f$_{sed}$) ranging from 2 (furthest right) to 5 (furthest left) at log g=5.0.  Also over-plotted are the clear (red short dashed lines) and partly cloudy (50\%) water + sulfide (blue long dashed) model predictions from \citet{Saumon12} and \citet{Morley14} respectively at five different gravities (log g=3.0, 3.5, 4.0, 4.5, 5.0).  All models shown are in chemical equilibrium at solar metallicity.  The corresponding T$_{eff}$ for the water cloud models are shown to the left of the tracks.  We show the 5$\sigma$ detection limit for W0855 (green five point star with right pointing arrow) as well as the 2.6$\sigma$ detection (filled triangle).}
\label{CMD_T0855}
\end{figure*}

The estimated position for W0855 on 14 May 2014 was deduced from the astrometry and NEOWISE-R position in the \citet{Wright14} paper taken just 10 days prior to the FourStar image. We estimated a 1 pixel uncertainty by applying the FourStar plate-scale to the \citet{Wright14} astrometric uncertainties added in quadrature with the WCS uncertainty (0.1$\arcsec$ based on the fit to 2MASS positions).  Aperture photometry was conducted at positions of x$\pm$1 pixel, y$\pm$1 pixel from the expected position. This accounts for the astrometric uncertainty and pinpoints the exact location of maximum flux associated with W0855.   The estimated position of W0855 using the propagated NEOWISE-R position is (RA=133.7861633, DEC=-7.2442513, EPOCH=56792.0) whereas the 2.6$\sigma$ position appears at (RA=133.7861696, DEC=-7.2442643, EPOCH=56792.0).  Therefore our detected FourStar position is $<$ 0.3 pixel from the expectation or consistent within the astrometric uncertainty. We note that the difference between the expected and measured position for W0855 is primarily in declination. The same procedure was used for the randomly chosen background pixels, therefore regardless of the 1 pixel buffer, the detections and consequent aperture comparisons are robust.  

The cumulative and probability distribution functions for a detection at the position of W0855 are shown in Figure ~\ref{Fig:CDF}.  We find that the W0855 aperture is brighter than $\sim$ 98\% of the 1000 random apertures or equivalent to a 2.6$\sigma$ detection.  The uncertainty for this low-signal source is assymetric and we measure $J3$=24.8$^{+0.53}_{-0.35}$ -- or J$_{MKO}$=25.0$^{+0.53}_{-0.35}$ when the photometric relations from \citealt{Tinney14} are applied.  This value is consistent with the 5$\sigma$ J$_{MKO}$ $>$ 24.0 limit on our photometry\footnote{Using gaussian statistics there should be a 1.16 mag difference between a 2.6$\sigma$ and 5$\sigma$ detection but we have rounded the upper limit magnitude}.

\section{WISE0855 on COLOR MAGNITUDE DIAGRAMS}\label{section:clouds} 
Color-Magnitude diagrams are a tool for investigating atmospheric properties of the brown dwarf population as well as testing model predictions (e.g. \citealt{Tinney03, Tinney14}, \citealt{Vrba04}, \citealt{Dupuy12}, \citealt{Faherty12, Faherty13}, \citealt{Patten06}, \citealt{Leggett10,Leggett13}). For warm brown dwarfs -- primarily L dwarfs -- iron and silicate grain clouds form thick dust layers and  largely shape observable properties (e.g. \citealt{Lodders03}, \citealt{Helling06}).  As objects cool into the T dwarf phase, L dwarf clouds rapidly clear and models predict that less refractory sulfide and salt clouds condense in the photospheres (e.g. \citealt{Lodders99}, \citealt{Morley12}).  For the coldest brown dwarfs, the Y dwarfs at T$_{eff}$  $<$ $\sim$450 K (see Figure ~\ref{T0855_SpT}), \citet{Morley14} predict that water condenses in the upper atmosphere to form ice clouds.  At T$_{eff}$  $<$ 350 - 375 K, \citet{Morley14} also show that these water clouds become optically thick, scatter light at optical wavelengths through $J$ band and absorb in the infrared with prominent features.  As such, infrared color-magnitude diagrams are particularly telling about the condensate species present in the photospheres of cold brown dwarfs.

Using our new photometry (both the 5$\sigma$ limit and 2.6$\sigma$ detection), we look for evidence of water clouds in W0855.  Figure~\ref{CMD_T0855} shows the J$_{MKO}$ $-$ W2  versus M$_{W2}$ for late-type T and Y dwarfs (\citealt{Dupuy13}, \citealt{Tinney12}, \citealt{Beichman14}, \citealt{Marsh13}).   As stated above, \citet{Morley14} recently published a set of atmosphere models applicable for the coldest brown dwarfs discovered (200 K$<$ T$_{eff}$ $<$ 450 K) that include the influence of water clouds in addition to the sulfide clouds described in \citet{Morley12}  (the model is a hybrid with 50\% cloudy and 50\% clear).   We have over plotted both of these model predictions on Figure~\ref{CMD_T0855}.  The sulfide cloud model predictions that include Na$_{2}$S, MnS, ZnS, Cr, and KCl condensate clouds and are applicable for objects in the range (400 K$<$ T$_{eff}$ $<$ 1300 K) are shown at log g=5.0 for differing sedimentation parameters (f$_{sed}$=5 to f$_{sed}$=2 as first defined in \citealt{Ackerman01}).  The water cloud models are shown at f$_{sed}$=5 for log $g$= 3.0 to log $g$=5.0 with the applicable T$_{eff}$ for the given position labeled on Figure~\ref{CMD_T0855}. The clear (or cloudless) model predictions from \citet{Saumon12} are also shown for log $g$=3.0 to log $g$=5.0, although those show little diversity for differing gravity values.  At T$_{eff}$ $<$ 350 K the clear  model predictions begin to diverge from those that include water clouds, giving a lever for differentiating between the two opacity realms.  We note that all models displayed are in chemical equilibrium for solar metallicity.  When the range of these parameters are fully incorporated into the models (e.g. non-solar metallicity, non-equilibrium chemistry),  predictions may vary (see for example \citealt{Burningham13}, \citealt{Hubeny07}).   

 The previous near infrared limit of W0855 of $J_{MKO}$ $>$ 23.0 was not sufficiently deep enough to discriminate between model predictions  (\citealt{Luhman14b}).  Our new 5$\sigma$ limit shifts the source 1 magnitude fainter and into the blue end of atmosphere model predictions that include water clouds.  The 2.6$\sigma$ detection of J$_{MKO}$=25.0$^{+0.53}_{-0.35}$ places W0855 2.7$\sigma$ from the clear model and squarely within sulfide and water cloud predictions for an object at 
225 $<$ T$_{eff}$$<$ 250 K.  The uncertainty is inclusive of all gravity predictions, however, given the large tangential velocity (v$_{tan}$ = 85$\pm$ 9 km$^{-1}$) of this source makes a young age unlikely (e.g. \citealt{Faherty09}). 

\section{CONCLUSIONS}
The recently discovered brown dwarf WISE J085510.83-071442.5 (W0855) is important for cold atmosphere studies and a gateway object for directly imaged gas-giant characterization programs.  In this work, we present a deep near-infrared $J3$ band image for the source. We find a  lower limit on its photometry of $J3$ $>$ 23.8 which we convert to J$_{MKO}$ $>$ 24.0 at 5$\sigma$ and a 2.6$\sigma$ detection of $J3$=24.8$^{+0.53}_{-0.35}$ which we convert to J$_{MKO}$=25.0$^{+0.53}_{-0.35}$.   With J$_{MKO}$ $-$ W2 = 10.984$^{+0.53}_{-0.35}$ at 2.6$\sigma$ -- or equivalently an upper limit of J$_{MKO}$ $-$ W2 $>$ 9.984 at 5$\sigma$ -- W0855 is the reddest brown dwarf yet characterized.   

Investigating the position of W0855 on the J$_{MKO}$ $-$ W2 versus M$_{W2}$ color-magnitude diagram reveals that this is the first compact source outside our solar system to have evidence for water ice clouds.  The 5$\sigma$ limit reported here-in is 1 magnitude deeper than previous works and places W0855 at the blue end of the water cloud model predictions from \citet{Morley14}.  The 2.6$\sigma$ detection places W0855 2.7$\sigma$ from the cloudless models and squarely along the water cloud model predictions for a source at 225K $<$T$_{eff}$ $<$ 250K.  A deeper near-infrared image coupled with advancements in the models to include non-equlibrium chemistry and varying metallicity at the temperature of W0855, will refine this result at a higher confidence level.

\acknowledgments{Acknowledgements }
This publication uses data gathered with the 6.5 meter Magellan Telescopes located at Las Campanas Observatory, Chile and we thank the operators J. Araya, and G. Martin and engineer D. Ossip for assistance in acquiring data.  The authors thank E. Wright and D. Saumon for useful comments on the manuscript.  Australian access to the Magellan Telescopes was supported through the National Collaborative Research Infrastructure and Collaborative Research Infrastructure Strategies of the Australian Federal Government.  This research was supported by Australian Research Council grants DP0774000 and DP130102695.
 This publication makes use of data products from WISE, which is a joint project of the University of California, Los Angeles, and the Jet Propulsion Laboratory (JPL)/California Institute of Technology (Caltech), funded by the National Aeronautics and Space Administration (NASA).  This research has made use of the NASA/ IPAC Infrared Science Archive, which is operated by the Jet Propulsion Laboratory, California Institute of Technology, under contract with the National Aeronautics and Space Administration.


\clearpage


\clearpage

\begin{thebibliography}{}
\bibitem[{{Ackerman} \& {Marley}(2001)}]{Ackerman01}
{Ackerman}, A.~S. \& {Marley}, M.~S. 2001, \apj, 556, 872

\bibitem[{{Beichman} {et~al.}(2014){Beichman}, {Gelino}, {Kirkpatrick},
  {Cushing}, {Dodson-Robinson}, {Marley}, {Morley}, \& {Wright}}]{Beichman14}
{Beichman}, C., {Gelino}, C.~R., {Kirkpatrick}, J.~D., {Cushing}, M.~C.,
  {Dodson-Robinson}, S., {Marley}, M.~S., {Morley}, C.~V., \& {Wright}, E.~L.
  2014, \apj, 783, 68

\bibitem[{{Burgasser} {et~al.}(2011){Burgasser}, {Cushing}, {Kirkpatrick},
  {Gelino}, {Griffith}, {Looper}, {Tinney}, {Simcoe}, {Bochanski}, {Skrutskie},
  {Mainzer}, {Thompson}, {Marsh}, {Bauer}, \& {Wright}}]{Burgasser11}
{Burgasser}, A.~J., {Cushing}, M.~C., {Kirkpatrick}, J.~D., {Gelino}, C.~R.,
  {Griffith}, R.~L., {Looper}, D.~L., {Tinney}, C., {Simcoe}, R.~A.,
  {Bochanski}, J.~J., {Skrutskie}, M.~F., {Mainzer}, A., {Thompson}, M.~A.,
  {Marsh}, K.~A., {Bauer}, J.~M., \& {Wright}, E.~L. 2011, \apj, 735, 116

\bibitem[{{Burningham} {et~al.}(2013){Burningham}, {Cardoso}, {Smith},
  {Leggett}, {Smart}, {Mann}, {Dhital}, {Lucas}, {Tinney}, {Pinfield}, {Zhang},
  {Morley}, {Saumon}, {Aller}, {Littlefair}, {Homeier}, {Lodieu}, {Deacon},
  {Marley}, {van Spaandonk}, {Baker}, {Allard}, {Andrei}, {Canty}, {Clarke},
  {Day-Jones}, {Dupuy}, {Fortney}, {Gomes}, {Ishii}, {Jones}, {Liu},
  {Magazz{\'u}}, {Marocco}, {Murray}, {Rojas-Ayala}, \&
  {Tamura}}]{Burningham13}
{Burningham}, B., {Cardoso}, C.~V., {Smith}, L., {Leggett}, S.~K., {Smart},
  R.~L., {Mann}, A.~W., {Dhital}, S., {Lucas}, P.~W., {Tinney}, C.~G.,
  {Pinfield}, D.~J., {Zhang}, Z., {Morley}, C., {Saumon}, D., {Aller}, K.,
  {Littlefair}, S.~P., {Homeier}, D., {Lodieu}, N., {Deacon}, N., {Marley},
  M.~S., {van Spaandonk}, L., {Baker}, D., {Allard}, F., {Andrei}, A.~H.,
  {Canty}, J., {Clarke}, J., {Day-Jones}, A.~C., {Dupuy}, T., {Fortney}, J.~J.,
  {Gomes}, J., {Ishii}, M., {Jones}, H.~R.~A., {Liu}, M., {Magazz{\'u}}, A.,
  {Marocco}, F., {Murray}, D.~N., {Rojas-Ayala}, B., \& {Tamura}, M. 2013,
  \mnras, 433, 457

\bibitem[{{Burningham} {et~al.}(2011{\natexlab{a}}){Burningham}, {Leggett},
  {Homeier}, {Saumon}, {Lucas}, {Pinfield}, {Tinney}, {Allard}, {Marley},
  {Jones}, {Murray}, {Ishii}, {Day-Jones}, {Gomes}, \& {Zhang}}]{Burningham11}
{Burningham}, B., {Leggett}, S.~K., {Homeier}, D., {Saumon}, D., {Lucas},
  P.~W., {Pinfield}, D.~J., {Tinney}, C.~G., {Allard}, F., {Marley}, M.~S.,
  {Jones}, H.~R.~A., {Murray}, D.~N., {Ishii}, M., {Day-Jones}, A., {Gomes},
  J., \& {Zhang}, Z.~H. 2011{\natexlab{a}}, \mnras, 414, 3590

\bibitem[{{Burningham} {et~al.}(2011{\natexlab{b}}){Burningham}, {Lucas},
  {Leggett}, {Smart}, {Baker}, {Pinfield}, {Tinney}, {Homeier}, {Allard},
  {Zhang}, {Gomes}, {Day-Jones}, {Jones}, {Kov{\'a}cs}, {Lodieu}, {Marocco},
  {Murray}, \& {Sip{\H o}cz}}]{Burningham11a}
{Burningham}, B., {Lucas}, P.~W., {Leggett}, S.~K., {Smart}, R., {Baker}, D.,
  {Pinfield}, D.~J., {Tinney}, C.~G., {Homeier}, D., {Allard}, F., {Zhang},
  Z.~H., {Gomes}, J., {Day-Jones}, A.~C., {Jones}, H.~R.~A., {Kov{\'a}cs}, G.,
  {Lodieu}, N., {Marocco}, F., {Murray}, D.~N., \& {Sip{\H o}cz}, B.
  2011{\natexlab{b}}, \mnras, 414, L90

\bibitem[{{Burrows} {et~al.}(1997){Burrows}, {Marley}, {Hubbard}, {Lunine},
  {Guillot}, {Saumon}, {Freedman}, {Sudarsky}, \& {Sharp}}]{Burrows97}
{Burrows}, A., {Marley}, M., {Hubbard}, W.~B., {Lunine}, J.~I., {Guillot}, T.,
  {Saumon}, D., {Freedman}, R., {Sudarsky}, D., \& {Sharp}, C. 1997, \apj, 491,
  856

\bibitem[{{Cushing} {et~al.}(2011){Cushing}, {Kirkpatrick}, {Gelino},
  {Griffith}, {Skrutskie}, {Mainzer}, {Marsh}, {Beichman}, {Burgasser},
  {Prato}, {Simcoe}, {Marley}, {Saumon}, {Freedman}, {Eisenhardt}, \&
  {Wright}}]{Cushing11}
{Cushing}, M.~C., {Kirkpatrick}, J.~D., {Gelino}, C.~R., {Griffith}, R.~L.,
  {Skrutskie}, M.~F., {Mainzer}, A., {Marsh}, K.~A., {Beichman}, C.~A.,
  {Burgasser}, A.~J., {Prato}, L.~A., {Simcoe}, R.~A., {Marley}, M.~S.,
  {Saumon}, D., {Freedman}, R.~S., {Eisenhardt}, P.~R., \& {Wright}, E.~L.
  2011, \apj, 743, 50

\bibitem[{{Cushing} {et~al.}(2014){Cushing}, {Kirkpatrick}, {Gelino}, {Mace},
  {Skrutskie}, \& {Gould}}]{Cushing14}
{Cushing}, M.~C., {Kirkpatrick}, J.~D., {Gelino}, C.~R., {Mace}, G.~N.,
  {Skrutskie}, M.~F., \& {Gould}, A. 2014, \aj, 147, 113

\bibitem[{{Cutri} {et~al.}(2003){Cutri}, {Skrutskie}, {van Dyk}, {Beichman},
  {Carpenter}, {Chester}, {Cambresy}, {Evans}, {Fowler}, {Gizis}, {Howard},
  {Huchra}, {Jarrett}, {Kopan}, {Kirkpatrick}, {Light}, {Marsh}, {McCallon},
  {Schneider}, {Stiening}, {Sykes}, {Weinberg}, {Wheaton}, {Wheelock}, \&
  {Zacarias}}]{Cutri03}
{Cutri}, R.~M., {Skrutskie}, M.~F., {van Dyk}, S., {Beichman}, C.~A.,
  {Carpenter}, J.~M., {Chester}, T., {Cambresy}, L., {Evans}, T., {Fowler}, J.,
  {Gizis}, J., {Howard}, E., {Huchra}, J., {Jarrett}, T., {Kopan}, E.~L.,
  {Kirkpatrick}, J.~D., {Light}, R.~M., {Marsh}, K.~A., {McCallon}, H.,
  {Schneider}, S., {Stiening}, R., {Sykes}, M., {Weinberg}, M., {Wheaton},
  W.~A., {Wheelock}, S., \& {Zacarias}, N. 2003, {2MASS All Sky Catalog of
  point sources.}, ed. {Cutri, R.~M., Skrutskie, M.~F., van Dyk, S., Beichman,
  C.~A., Carpenter, J.~M., Chester, T., Cambresy, L., Evans, T., Fowler, J.,
  Gizis, J., Howard, E., Huchra, J., Jarrett, T., Kopan, E.~L., Kirkpatrick,
  J.~D., Light, R.~M., Marsh, K.~A., McCallon, H., Schneider, S., Stiening, R.,
  Sykes, M., Weinberg, M., Wheaton, W.~A., Wheelock, S., \& Zacarias, N.}

\bibitem[{{Delorme} {et~al.}(2012){Delorme}, {Gagn{\'e}}, {Malo}, {Reyl{\'e}},
  {Artigau}, {Albert}, {Forveille}, {Delfosse}, {Allard}, \&
  {Homeier}}]{Delorme12}
{Delorme}, P., {Gagn{\'e}}, J., {Malo}, L., {Reyl{\'e}}, C., {Artigau}, E.,
  {Albert}, L., {Forveille}, T., {Delfosse}, X., {Allard}, F., \& {Homeier}, D.
  2012, \aap, 548, A26

\bibitem[{{Dupuy} \& {Kraus}(2013)}]{Dupuy13}
{Dupuy}, T.~J. \& {Kraus}, A.~L. 2013, Science, 341, 1492

\bibitem[{{Dupuy} \& {Liu}(2012)}]{Dupuy12}
{Dupuy}, T.~J. \& {Liu}, M.~C. 2012, \apjs, 201, 19

\bibitem[{{Faherty} {et~al.}(2009){Faherty}, {Burgasser}, {Cruz}, {Shara},
  {Walter}, \& {Gelino}}]{Faherty09}
{Faherty}, J.~K., {Burgasser}, A.~J., {Cruz}, K.~L., {Shara}, M.~M., {Walter},
  F.~M., \& {Gelino}, C.~R. 2009, \aj, 137, 1

\bibitem[{{Faherty} {et~al.}(2012){Faherty}, {Burgasser}, {Walter}, {Van der
  Bliek}, {Shara}, {Cruz}, {West}, {Vrba}, \& {Anglada-Escud{\'e}}}]{Faherty12}
{Faherty}, J.~K., {Burgasser}, A.~J., {Walter}, F.~M., {Van der Bliek}, N.,
  {Shara}, M.~M., {Cruz}, K.~L., {West}, A.~A., {Vrba}, F.~J., \&
  {Anglada-Escud{\'e}}, G. 2012, \apj, 752, 56

\bibitem[{{Faherty} {et~al.}(2013{\natexlab{a}}){Faherty}, {Cruz}, {Rice}, \&
  {Riedel}}]{Faherty13a}
{Faherty}, J.~K., {Cruz}, K.~L., {Rice}, E.~L., \& {Riedel}, A.
  2013{\natexlab{a}}, ArXiv e-prints

\bibitem[{{Faherty} {et~al.}(2013{\natexlab{b}}){Faherty}, {Rice}, {Cruz},
  {Mamajek}, \& {N{\'u}{\~n}ez}}]{Faherty13}
{Faherty}, J.~K., {Rice}, E.~L., {Cruz}, K.~L., {Mamajek}, E.~E., \&
  {N{\'u}{\~n}ez}, A. 2013{\natexlab{b}}, \aj, 145, 2

\bibitem[{{Helling} \& {Woitke}(2006)}]{Helling06}
{Helling}, C. \& {Woitke}, P. 2006, \aap, 455, 325

\bibitem[{{Hubeny} \& {Burrows}(2007)}]{Hubeny07}
{Hubeny}, I. \& {Burrows}, A. 2007, \apj, 669, 1248

\bibitem[{{Kirkpatrick} {et~al.}(2013){Kirkpatrick}, {Cushing}, {Gelino},
  {Beichman}, {Tinney}, {Faherty}, {Schneider}, \& {Mace}}]{Kirkpatrick13}
{Kirkpatrick}, J.~D., {Cushing}, M.~C., {Gelino}, C.~R., {Beichman}, C.~A.,
  {Tinney}, C.~G., {Faherty}, J.~K., {Schneider}, A., \& {Mace}, G.~N. 2013,
  \apj, 776, 128

\bibitem[{{Kirkpatrick} {et~al.}(2011){Kirkpatrick}, {Cushing}, {Gelino},
  {Griffith}, {Skrutskie}, {Marsh}, {Wright}, {Mainzer}, {Eisenhardt},
  {McLean}, {Thompson}, {Bauer}, {Benford}, {Bridge}, {Lake}, {Petty},
  {Stanford}, {Tsai}, {Bailey}, {Beichman}, {Bloom}, {Bochanski}, {Burgasser},
  {Capak}, {Cruz}, {Hinz}, {Kartaltepe}, {Knox}, {Manohar}, {Masters},
  {Morales-Calder{\'o}n}, {Prato}, {Rodigas}, {Salvato}, {Schurr}, {Scoville},
  {Simcoe}, {Stapelfeldt}, {Stern}, {Stock}, \& {Vacca}}]{Kirkpatrick11}
{Kirkpatrick}, J.~D., {Cushing}, M.~C., {Gelino}, C.~R., {Griffith}, R.~L.,
  {Skrutskie}, M.~F., {Marsh}, K.~A., {Wright}, E.~L., {Mainzer}, A.,
  {Eisenhardt}, P.~R., {McLean}, I.~S., {Thompson}, M.~A., {Bauer}, J.~M.,
  {Benford}, D.~J., {Bridge}, C.~R., {Lake}, S.~E., {Petty}, S.~M., {Stanford},
  S.~A., {Tsai}, C.-W., {Bailey}, V., {Beichman}, C.~A., {Bloom}, J.~S.,
  {Bochanski}, J.~J., {Burgasser}, A.~J., {Capak}, P.~L., {Cruz}, K.~L.,
  {Hinz}, P.~M., {Kartaltepe}, J.~S., {Knox}, R.~P., {Manohar}, S., {Masters},
  D., {Morales-Calder{\'o}n}, M., {Prato}, L.~A., {Rodigas}, T.~J., {Salvato},
  M., {Schurr}, S.~D., {Scoville}, N.~Z., {Simcoe}, R.~A., {Stapelfeldt},
  K.~R., {Stern}, D., {Stock}, N.~D., \& {Vacca}, W.~D. 2011, \apjs, 197, 19

\bibitem[{{Kirkpatrick} {et~al.}(2012){Kirkpatrick}, {Gelino}, {Cushing},
  {Mace}, {Griffith}, {Skrutskie}, {Marsh}, {Wright}, {Eisenhardt}, {McLean},
  {Mainzer}, {Burgasser}, {Tinney}, {Parker}, \& {Salter}}]{Kirkpatrick12}
{Kirkpatrick}, J.~D., {Gelino}, C.~R., {Cushing}, M.~C., {Mace}, G.~N.,
  {Griffith}, R.~L., {Skrutskie}, M.~F., {Marsh}, K.~A., {Wright}, E.~L.,
  {Eisenhardt}, P.~R., {McLean}, I.~S., {Mainzer}, A.~K., {Burgasser}, A.~J.,
  {Tinney}, C.~G., {Parker}, S., \& {Salter}, G. 2012, \apj, 753, 156

\bibitem[{{Kirkpatrick} {et~al.}(2014){Kirkpatrick}, {Schneider},
  {Fajardo-Acosta}, {Gelino}, {Mace}, {Wright}, {Logsdon}, {McLean}, {Cushing},
  {Skrutskie}, {Eisenhardt}, {Stern}, {Balokovi{\'c}}, {Burgasser}, {Faherty},
  {Lansbury}, {Rich}, {Skrzypek}, {Fowler}, {Cutri}, {Masci}, {Conrow},
  {Grillmair}, {McCallon}, {Beichman}, \& {Marsh}}]{Kirkpatrick14}
{Kirkpatrick}, J.~D., {Schneider}, A., {Fajardo-Acosta}, S., {Gelino}, C.~R.,
  {Mace}, G.~N., {Wright}, E.~L., {Logsdon}, S.~E., {McLean}, I.~S., {Cushing},
  M.~C., {Skrutskie}, M.~F., {Eisenhardt}, P.~R., {Stern}, D., {Balokovi{\'c}},
  M., {Burgasser}, A.~J., {Faherty}, J.~K., {Lansbury}, G.~B., {Rich}, J.~A.,
  {Skrzypek}, N., {Fowler}, J.~W., {Cutri}, R.~M., {Masci}, F.~J., {Conrow},
  T., {Grillmair}, C.~J., {McCallon}, H.~L., {Beichman}, C.~A., \& {Marsh},
  K.~A. 2014, \apj, 783, 122

\bibitem[{{Leggett} {et~al.}(2010){Leggett}, {Burningham}, {Saumon}, {Marley},
  {Warren}, {Smart}, {Jones}, {Lucas}, {Pinfield}, \& {Tamura}}]{Leggett10}
{Leggett}, S.~K., {Burningham}, B., {Saumon}, D., {Marley}, M.~S., {Warren},
  S.~J., {Smart}, R.~L., {Jones}, H.~R.~A., {Lucas}, P.~W., {Pinfield}, D.~J.,
  \& {Tamura}, M. 2010, \apj, 710, 1627

\bibitem[{{Leggett} {et~al.}(2013){Leggett}, {Morley}, {Marley}, {Saumon},
  {Fortney}, \& {Visscher}}]{Leggett13}
{Leggett}, S.~K., {Morley}, C.~V., {Marley}, M.~S., {Saumon}, D., {Fortney},
  J.~J., \& {Visscher}, C. 2013, \apj, 763, 130

\bibitem[{{Liu} {et~al.}(2013){Liu}, {Magnier}, {Deacon}, {Allers}, {Dupuy},
  {Kotson}, {Aller}, {Burgett}, {Chambers}, {Draper}, {Hodapp}, {Jedicke},
  {Kaiser}, {Kudritzki}, {Metcalfe}, {Morgan}, {Price}, {Tonry}, \&
  {Wainscoat}}]{Liu13}
{Liu}, M.~C., {Magnier}, E.~A., {Deacon}, N.~R., {Allers}, K.~N., {Dupuy},
  T.~J., {Kotson}, M.~C., {Aller}, K.~M., {Burgett}, W.~S., {Chambers}, K.~C.,
  {Draper}, P.~W., {Hodapp}, K.~W., {Jedicke}, R., {Kaiser}, N., {Kudritzki},
  R.-P., {Metcalfe}, N., {Morgan}, J.~S., {Price}, P.~A., {Tonry}, J.~L., \&
  {Wainscoat}, R.~J. 2013, \apjl, 777, L20

\bibitem[{{Lodders}(1999)}]{Lodders99}
{Lodders}, K. 1999, \apj, 519, 793

\bibitem[{{Lodders}(2003)}]{Lodders03}
---. 2003, \apj, 591, 1220

\bibitem[{{Lucas} {et~al.}(2011){Lucas}, {Tinney}, {Burningham}, {Leggett},
  {Pinfield}, {Smart}, {Jones}, {Marocco}, {Barber}, {Yurchenko}, {Tennyson},
  {Ishii}, {Tamura}, {Day-Jones}, {Adamson}, {Allard}, \& {Homeier}}]{Lucas11}
{Lucas}, P.~W., {Tinney}, C.~G., {Burningham}, B., {Leggett}, S.~K.,
  {Pinfield}, D.~J., {Smart}, R., {Jones}, H.~R.~A., {Marocco}, F., {Barber},
  R.~J., {Yurchenko}, S.~N., {Tennyson}, J., {Ishii}, M., {Tamura}, M.,
  {Day-Jones}, A.~C., {Adamson}, A., {Allard}, F., \& {Homeier}, D. 2011, in
  Astronomical Society of the Pacific Conference Series, Vol. 448, 16th
  Cambridge Workshop on Cool Stars, Stellar Systems, and the Sun, ed.
  C.~{Johns-Krull}, M.~K. {Browning}, \& A.~A. {West}, 339

\bibitem[{{Luhman}(2014\natexlab{a})}]{Luhman14a}
{Luhman}, K.~L. 2014\natexlab{a}, \apj, 781, 4

\bibitem[{Luhman(2014\natexlab{b})}]{Luhman14b}
Luhman, K.~L. 2014\natexlab{b}, The Astrophysical Journal, 786, L18

\bibitem[{{Marsh} {et~al.}(2013){Marsh}, {Wright}, {Kirkpatrick}, {Gelino},
  {Cushing}, {Griffith}, {Skrutskie}, \& {Eisenhardt}}]{Marsh13}
{Marsh}, K.~A., {Wright}, E.~L., {Kirkpatrick}, J.~D., {Gelino}, C.~R.,
  {Cushing}, M.~C., {Griffith}, R.~L., {Skrutskie}, M.~F., \& {Eisenhardt},
  P.~R. 2013, \apj, 762, 119

\bibitem[{{Morley} {et~al.}(2012){Morley}, {Fortney}, {Marley}, {Visscher},
  {Saumon}, \& {Leggett}}]{Morley12}
{Morley}, C.~V., {Fortney}, J.~J., {Marley}, M.~S., {Visscher}, C., {Saumon},
  D., \& {Leggett}, S.~K. 2012, \apj, 756, 172

\bibitem[{{Morley} {et~al.}(2014){Morley}, {Marley}, {Fortney}, {Lupu},
  {Saumon}, {Greene}, \& {Lodders}}]{Morley14}
{Morley}, C.~V., {Marley}, M.~S., {Fortney}, J.~J., {Lupu}, R., {Saumon}, D.,
  {Greene}, T., \& {Lodders}, K. 2014, \apj, 787, 78

\bibitem[{{Patten} {et~al.}(2006){Patten}, {Stauffer}, {Burrows}, {Marengo},
  {Hora}, {Luhman}, {Sonnett}, {Henry}, {Raghavan}, {Megeath}, {Liebert}, \&
  {Fazio}}]{Patten06}
{Patten}, B.~M., {Stauffer}, J.~R., {Burrows}, A., {Marengo}, M., {Hora},
  J.~L., {Luhman}, K.~L., {Sonnett}, S.~M., {Henry}, T.~J., {Raghavan}, D.,
  {Megeath}, S.~T., {Liebert}, J., \& {Fazio}, G.~G. 2006, \apj, 651, 502

\bibitem[{{Persson} {et~al.}(2013){Persson}, {Murphy}, {Smee}, {Birk},
  {Monson}, {Uomoto}, {Koch}, {Shectman}, {Barkhouser}, {Orndorff}, {Hammond},
  {Harding}, {Scharfstein}, {Kelson}, {Marshall}, \& {McCarthy}}]{Persson13}
{Persson}, S.~E., {Murphy}, D.~C., {Smee}, S., {Birk}, C., {Monson}, A.~J.,
  {Uomoto}, A., {Koch}, E., {Shectman}, S., {Barkhouser}, R., {Orndorff}, J.,
  {Hammond}, R., {Harding}, A., {Scharfstein}, G., {Kelson}, D., {Marshall},
  J., \& {McCarthy}, P.~J. 2013, \pasp, 125, 654

\bibitem[{{Saumon} {et~al.}(2012){Saumon}, {Marley}, {Abel}, {Frommhold}, \&
  {Freedman}}]{Saumon12}
{Saumon}, D., {Marley}, M.~S., {Abel}, M., {Frommhold}, L., \& {Freedman},
  R.~S. 2012, \apj, 750, 74

\bibitem[{{Tinney} {et~al.}(2003){Tinney}, {Burgasser}, \&
  {Kirkpatrick}}]{Tinney03}
{Tinney}, C.~G., {Burgasser}, A.~J., \& {Kirkpatrick}, J.~D. 2003, \aj, 126,
  975

\bibitem[{{Tinney} {et~al.}(2014){Tinney}, {Faherty}, {Kirkpatrick}, {Cushing},
  {Morley}, \& {Wright}}]{Tinney14}
{Tinney}, C.~G., {Faherty}, J.~K., {Kirkpatrick}, J.~D., {Cushing}, M.~C.,
  {Morley}, C.~V., \& {Wright}, E.~L. 2014, \apj, submitted

\bibitem[{{Tinney} {et~al.}(2012){Tinney}, {Faherty}, {Kirkpatrick}, {Wright},
  {Gelino}, {Cushing}, {Griffith}, \& {Salter}}]{Tinney12}
{Tinney}, C.~G., {Faherty}, J.~K., {Kirkpatrick}, J.~D., {Wright}, E.~L.,
  {Gelino}, C.~R., {Cushing}, M.~C., {Griffith}, R.~L., \& {Salter}, G. 2012,
  \apj, 759, 60

\bibitem[{{Vrba} {et~al.}(2004){Vrba}, {Henden}, {Luginbuhl}, {Guetter},
  {Munn}, {Canzian}, {Burgasser}, {Kirkpatrick}, {Fan}, {Geballe},
  {Golimowski}, {Knapp}, {Leggett}, {Schneider}, \& {Brinkmann}}]{Vrba04}
{Vrba}, F.~J., {Henden}, A.~A., {Luginbuhl}, C.~B., {Guetter}, H.~H., {Munn},
  J.~A., {Canzian}, B., {Burgasser}, A.~J., {Kirkpatrick}, J.~D., {Fan}, X.,
  {Geballe}, T.~R., {Golimowski}, D.~A., {Knapp}, G.~R., {Leggett}, S.~K.,
  {Schneider}, D.~P., \& {Brinkmann}, J. 2004, \aj, 127, 2948

\bibitem[{{Wright} {et~al.}(2010){Wright}, {Eisenhardt}, {Mainzer}, {Ressler},
  {Cutri}, {Jarrett}, {Kirkpatrick}, {Padgett}, {McMillan}, {Skrutskie},
  {Stanford}, {Cohen}, {Walker}, {Mather}, {Leisawitz}, {Gautier}, {McLean},
  {Benford}, {Lonsdale}, {Blain}, {Mendez}, {Irace}, {Duval}, {Liu}, {Royer},
  {Heinrichsen}, {Howard}, {Shannon}, {Kendall}, {Walsh}, {Larsen}, {Cardon},
  {Schick}, {Schwalm}, {Abid}, {Fabinsky}, {Naes}, \& {Tsai}}]{Wright10}
{Wright}, E.~L., {Eisenhardt}, P.~R.~M., {Mainzer}, A.~K., {Ressler}, M.~E.,
  {Cutri}, R.~M., {Jarrett}, T., {Kirkpatrick}, J.~D., {Padgett}, D.,
  {McMillan}, R.~S., {Skrutskie}, M., {Stanford}, S.~A., {Cohen}, M., {Walker},
  R.~G., {Mather}, J.~C., {Leisawitz}, D., {Gautier}, III, T.~N., {McLean}, I.,
  {Benford}, D., {Lonsdale}, C.~J., {Blain}, A., {Mendez}, B., {Irace}, W.~R.,
  {Duval}, V., {Liu}, F., {Royer}, D., {Heinrichsen}, I., {Howard}, J.,
  {Shannon}, M., {Kendall}, M., {Walsh}, A.~L., {Larsen}, M., {Cardon}, J.~G.,
  {Schick}, S., {Schwalm}, M., {Abid}, M., {Fabinsky}, B., {Naes}, L., \&
  {Tsai}, C.-W. 2010, \aj, 140, 1868

\bibitem[{{Wright} {et~al.}(2014){Wright}, {Mainzer}, {Kirkpatrick}, {Masci},
  {Cushing}, {Bauer}, {Fajardo-Acosta}, {Gelino}, {Beichman}, {Skrutskie},
  {Grav}, {Eisenhardt}, \& {Cutri}}]{Wright14}
{Wright}, E.~L., {Mainzer}, A., {Kirkpatrick}, J.~D., {Masci}, F., {Cushing},
  M.~C., {Bauer}, J., {Fajardo-Acosta}, S., {Gelino}, C.~R., {Beichman}, C.~A.,
  {Skrutskie}, M.~F., {Grav}, T., {Eisenhardt}, P.~R.~M., \& {Cutri}, R. 2014,
  ArXiv e-prints

\end{thebibliography}
\end{document}